\documentclass[twocolumn,amsmath,showpacs,aps,prl]{revtex4-1}

\usepackage{graphicx}

\begin{document}

\title{Graphene valley filter using a line defect}

\author{D.~Gunlycke}
\affiliation{Naval Research Laboratory, Washington, D.C.~20375, USA}
\author{C.~T.~White}
\affiliation{Naval Research Laboratory, Washington, D.C.~20375, USA}

\begin{abstract}
With its two degenerate valleys at the Fermi level, the band structure of graphene provides the opportunity to develop unconventional electronic applications.  Herein, we show that electron and hole quasiparticles in graphene can be filtered according to which valley they occupy without the need to introduce confinement.  The proposed valley filter is based on scattering off a recently observed line defect in graphene.  Quantum transport calculations show that the line defect is semitransparent and that quasiparticles arriving at the line defect with a high angle of incidence are transmitted with a valley polarization near 100\%.
\end{abstract}


\maketitle

%===========================================================================%

Owing to its exceptional electron\cite{Bolo08_1} and thermal\cite{Bala08_1} transport properties, graphene\cite{Novo04_1} is a promising material for use in advanced low-energy electronic applications.  As graphene is a semimetal with no band gap,\cite{Wall47_1} it cannot in its intrinsic form be used as a replacement material for silicon and other semiconductors in conventional electronics.  However, with its bands at the Fermi level defining two conical valleys, graphene might instead offer novel electronic applications.  Before such applications can be realized, the control over transport in graphene needs to be improved.  Understanding the effects a recently observed line defect\cite{Lahi10} and grain boundaries\cite{Huan11} have on the transport properties might be the key.  While there are several theoretical studies that have investigated the latter grain boundaries,\cite{Malo10,Yazy10,Liu10,Yazy10a,Huan11} the line defect has so far remained relatively unexplored, even though the line defect is always straight and its adjoining grains are aligned, arguably making the line defect a more suitable structure for controlled transport in graphene.

\begin{figure}
    \includegraphics{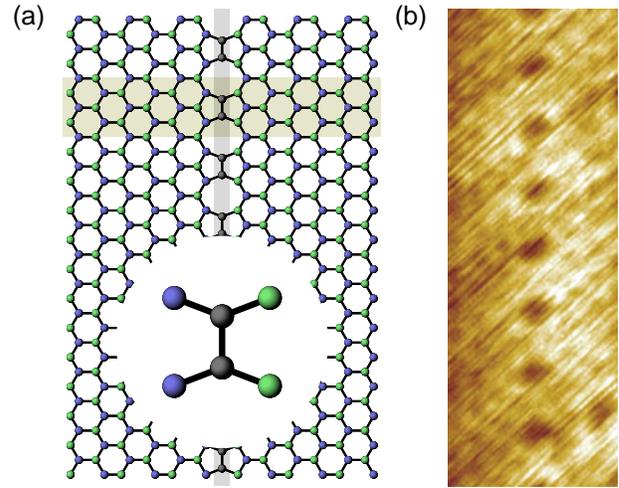}
    \caption{Extended line defect in graphene.  (a) Arrangement of the carbon atoms around the line defect highlighted in gray.  The structure exhibits translational symmetry along the defect with a primitive cell shown in beige.  The two sublattices in graphene indicated by blue and green atoms reverse upon reflection at the line defect, as can be clearly seen in the enlarged overlay.  (b) Scanning tunneling microscope image of the line defect in graphene (Image adapted from Ref.\,[\onlinecite{Lahi10}]).}
    \label{f.1}
\end{figure}

This Letter investigates the transport properties of the atomically precise, self-assembled graphene line defect shown in Fig.\,\ref{f.1} and shows that this line defect is semitransparent and can be used as a valley filter.
Electron and hole quasiparticles can either transmit through the line defect without changing direction or reflect following the law of specular reflection.  The transmission and reflection probabilities depend on the valley degree of freedom, thus allowing the quasiparticles to be filtered according to their valley degree of freedom with a polarization near 100\% for quasiparticles arriving at the line defect with a high angle of incidence.  This filter is much different from the valley filter proposed by Rycerz {\it et al.},\cite{Ryce07} which relies on the isolation of a few one-dimensional channels in a narrowly confined region.  The latter filter has not yet been demonstrated, presumably due to challenges in fabricating the structure, which requires a sub-10\,nm constriction with saturated zigzag edges.  In contrast, the filter proposed herein relies on the two-dimensional geometry of graphene and its required structure has already been observed,\cite{Lahi10} as demonstrated by the micrograph in Fig.\,\ref{f.1}.  The valley filter is expected to be a central component in valleytronics,\cite{Ryce07} just as the spin filter is central in spintronics.  Electronics that make use of the two valleys in graphene is attractive because the valleys are separated by a large wave vector, making valley information robust against scattering from slowly varying potentials,\cite{Ando98} including scattering caused by intravalley acoustic phonons that often limit coherent low-bias devices to low-temperature operation.  Therefore, the valley information generated by the filter proposed herein could in principle be preserved even in a diffusive charge transport regime.

\begin{figure}
    \includegraphics{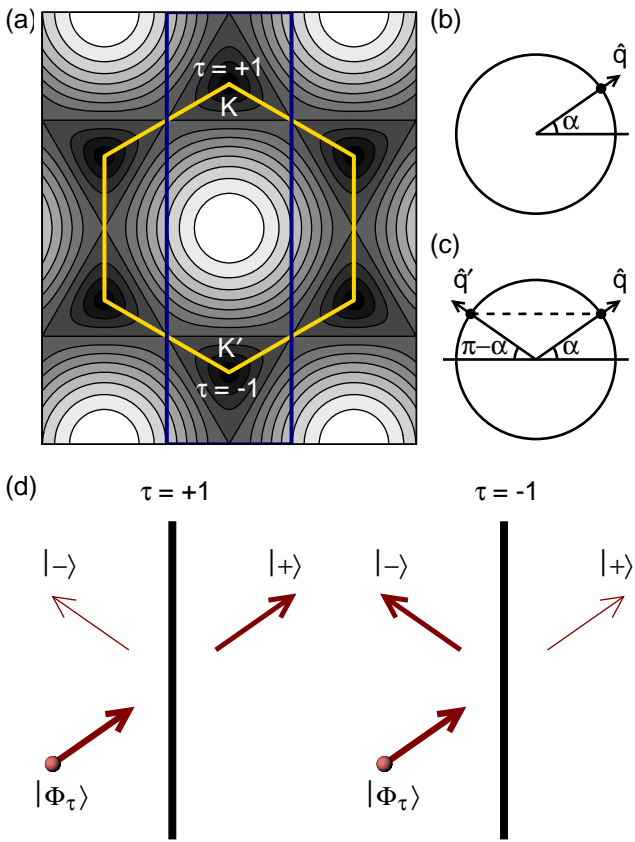}
    \caption{Valley states scattering off the line defect.  (a) Energies close to the Fermi level in graphene have dark contours and are located in the corners of the first Brillouin zone contained within the gold hexagon.  The two valleys, K and K$'$, are identified as the two disjoint low-energy regions in the reciprocal primitive cell enclosed by the blue rectangle.  (b) An incident quasiparticle state is defined by the valley index $\tau$ and wave vector $\vec{q}$, where the latter points in the direction $\hat{q}$ given by the angle of incidence $\alpha$.  (c) Owing to energy and momentum conservation along the line defect, there are only two nonevanescent scattered states allowed.  (d) The sublattice symmetric $|+\rangle$ and antisymmetric $|-\rangle$ components of the incident state $|\Phi_\tau\rangle$ are transmitted and reflected, respectively.  The thickness of each arrow indicates the probability the quasiparticle will follow the respective path.}
    \label{f.2}
\end{figure}

Consider a low-energy electron (hole) quasi\-particle with energy $\varepsilon$ and valley index $\tau$ approaching the line defect from the left (right) at the angle of incidence $\alpha$.  Asymptotically far from the line defect, the quasi\-particle occupies a graphene state $|\Phi_\tau\rangle$, where $\tau=\pm1$ is a valley index.  See Fig.\,\ref{f.2}.  Let the quasi\-particle wave vector $\vec{q}=(q_x,q_y)$ be the wave vector measured from the center of the occupied valley, located at $\vec{K}_\tau=4\pi\tau\hat{y}/3a$, where $a$ is the graphene lattice constant and $\hat{y}$ is the unit vector along the line defect.  To first order in $q\equiv|\vec{q}|$, the nearest-neighbor tight-binding Hamiltonian in graphene\cite{Wall47_1} can be expressed\cite{Slon57} as $H_\tau=\hbar v_F\left(q_x\sigma_y+\tau q_y\sigma_x\right)$, where $v_F=\sqrt{3}|\gamma|a/2\hbar$ is the Fermi velocity with the nearest-neighbor hopping parameter $\gamma\approx-2.6$\,eV, and $\sigma_x$ and $\sigma_y$ are Pauli matrices.  The Hamiltonian has energy eigenvalues $E=\eta\varepsilon$, where $\varepsilon=\hbar v_Fq$ and $\eta=+1$ ($-1$) if the quasi\-particle is an electron (hole).  From the quasi\-particle energy dispersion $\varepsilon$, it follows that the quasi\-particle group velocity is $v_F\hat{q}$, where $\hat{q}$ is the unit vector in the direction $\vec{q}$.  Because the quasi\-particle travels in the direction $\vec{q}$, $q_x=\eta q\cos\alpha$ and $q_y=\eta q\sin\alpha$.  Using these relations, the eigenstate of the graphene Hamiltonian for a given $\tau$ and $\alpha$ can be expressed as
\begin{equation}
	|\Phi_\tau\rangle=\frac{1}{\sqrt{2}}\left(|A\rangle+ie^{-i\tau\alpha}|B\rangle\right),
	\label{e.1}
\end{equation}
where $|A\rangle$ and $|B\rangle$ refer to the two sublattices in graphene.

The structure in Fig.\,\ref{f.1} exhibits a useful symmetry line through the line defect.  In the limit $q\rightarrow0$, the reflection operator commutes with the graphene translation operator perpendicular to the line defect.  Therefore, symmetry-adapted states $|\pm\rangle$ can be constructed that are simultaneous eigenstates of the graphene Hamiltonian and the reflection operator.  As the reflection operator maps $A$ sites onto $B$ sites, and vice versa, it can be represented by the operator $\sigma_x$ acting on the two sublattices.  From the eigenstates of $\sigma_x$, one obtains
\begin{equation}
	|\pm\rangle=\frac{1}{\sqrt{2}}\left(|A\rangle\pm|B\rangle\right).
	\label{e.2}
\end{equation}
The graphene state \eqref{e.1} expressed in the symmetry-adapted basis is
\begin{equation}
	|\Phi_\tau\rangle=\frac{1+ie^{-i\tau\alpha}}{2}|+\rangle+\frac{1-ie^{-i\tau\alpha}}{2}|-\rangle.
	\label{e.3}
\end{equation}

The full Hamiltonian describing the system can be divided into three terms, $\mathcal{H}_\tau = H_\tau + H_D + V$, which represent graphene, the isolated line defect, and the interaction between graphene and the line defect, respectively.  As each term commutes with the reflection operator, the full Hamiltonian must commute with the reflection operator, and thus, the eigenstates of $\mathcal{H}_\tau$ in the symmetry-adapted basis are either symmetric or antisymmetric about the line defect.  Antisymmetric states have a node at the line defect, and as a result, there are no matrix elements within the nearest-neighbor model coupling the left and right sides.  Therefore, antisymmetric states cannot contribute to any transmission across the line defect.  As shown below, however, there are two symmetric states at the Fermi level without a node on the line defect.  As these states are extended eigenstates of the full Hamiltonian, they carry quasi\-particles across the line defect without scattering.  Thus, we can conclude that the transmission probability of the quasi\-particle approaching the line defect is
\begin{equation}
    T_\tau = \left|\langle+|\Phi_\tau\rangle\right|^2=\frac{1}{2}\left(1+\tau\sin\alpha\right).
    \label{e.4}
\end{equation}

\begin{figure}
    \includegraphics{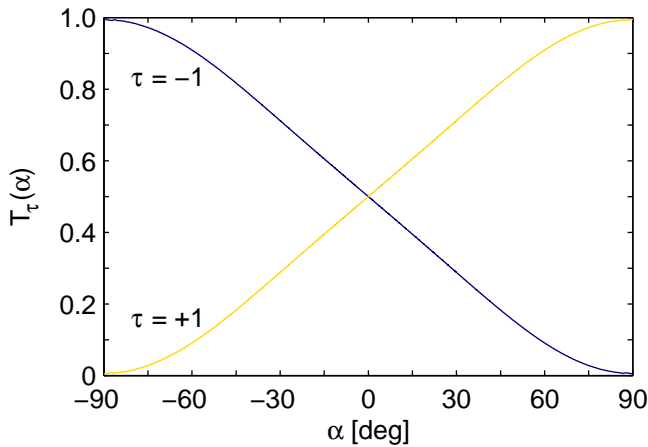}
    \caption{The probability that an incident quasi\-particle at the Fermi level with valley index $\tau$ and angle of incidence $\alpha$ will transmit through the line defect.}
    \label{f.3}
\end{figure}

As the sum of the transmission probabilities in Eq.\,\eqref{e.4} over the two valleys is $\sum_\tau T_\tau=1$, we can also conclude that the line defect is semitransparent.  The semitransparency follows from the relation $\langle+|\Phi_{-\tau}\rangle=\langle-|\Phi_\tau\rangle^*$ and the normalization of $|\Phi_\tau\rangle$.  See Fig.\,\ref{f.2}.  Fig.\,\ref{f.3} shows that the transmission probability of a quasi\-particle varies significantly with its angle of incidence $\alpha$.
At a high angle of incidence, there is almost full transmission or reflection, depending on the valley index $\tau$.

Owing to the semitransparency, given an unpolarized beam of incident quasi\-particles, the probability $P_\tau$ that a transmitted quasi\-particle has valley index $\tau$ is given by $T_\tau$.  The polarization $\mathcal{P}\equiv\langle\tau\rangle=P_{+1}-P_{-1}$ of the transmitted beam is
\begin{equation}
    \label{e.5}
    \mathcal{P} = \sin\alpha.
\end{equation}
This expression shows that an unpolarized beam of quasi\-particles approaching the line defect at a high angle of incidence will lead to outgoing transmitted and reflected beams that are almost completely polarized.

\begin{figure}
    \includegraphics{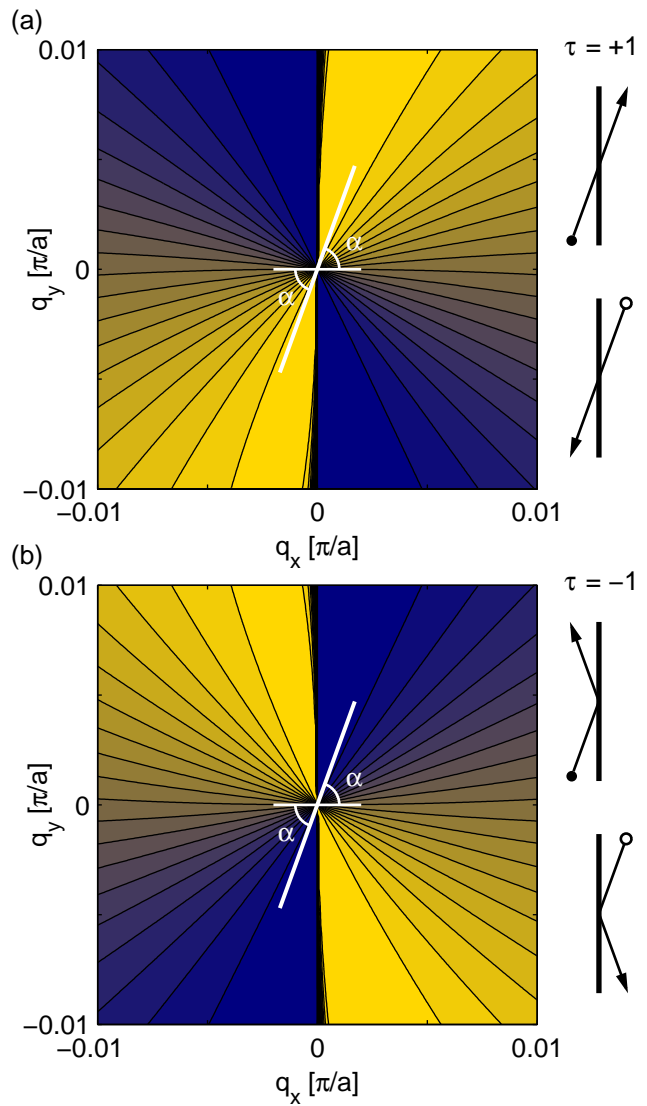}
    \caption{Transmission probability for a quasiparticle with finite energy.  (a) The transmission probability for a quasi\-particle approaching the line defect at an angle of incidence $\alpha$ is indicated by the brightness, ranging from 0 to 1 in increments of 0.05.  For the angle $\alpha$ shown, the transmission is near one if the quasiparticle has valley index $\tau=+1$, as illustrated on the right, where the filled (open) circle represents an electron (hole) quasiparticle.  (b) The transmission probability for the corresponding quasiparticle with valley index $\tau=-1$.}
    \label{f.4}
\end{figure}

The results presented above are based on symmetry arguments that are valid only in the limit as the quasi\-particle energy $\varepsilon\rightarrow 0$.  The results, however, hold to an excellent approximation as long as $\varepsilon\ll\hbar v_F/a\approx2.3$\,eV.  To show this, we have performed numerical transport calculations that are treated exactly within the nearest-neighbor tight-binding model.  The transmission probability is shown in Fig.\,\ref{f.4} as a function of the components of the wave vector $\vec{q}$.
Tracing the transmission probability in the figure along any constant energy contour, which is almost perfectly circular due to the approximately conic dispersion, yields a dependence on the angle of incidence that is virtually indistinguishable from that in Fig.\,\ref{f.3}, thus confirming that almost full polarization can be achieved as long as $qa\ll1$ or $\varepsilon\ll\hbar v_F/a$.  To further test the robustness of the valley filter, we have performed calculations with interactions across the line defect and with potentials on the line defect and their neighboring sites.  None of these tests led to results qualitatively different from those presented in Fig.\,\ref{f.4}.

Because the structure in Fig.\,\ref{f.1}(a) exhibits a symmetry plane through the center of the primitive cell, there is time-reversal symmetry in the direction along the line defect.  This time-reversal symmetry implies that the transmission probability of a quasi\-particle with valley index $-\tau$ can be obtained from that of a quasi\-particle with valley index $\tau$ by letting $q_y\rightarrow -q_y$.  This relationship between $\tau=\pm1$ can be seen in Fig.\,\ref{f.4}.  Note, however, that the transmission probabilities in Fig.\,\ref{f.4} are not symmetric about $q_y$, and thus one can conclude that the scattering of a quasi\-particle depends on the valley index, which is a necessary requirement for a valley filter.  As both graphene and the line defect have Hamiltonians exhibiting electron--hole symmetry, one might expect that $T_\tau(\vec{q}) = T_\tau(-\vec{q})$.  The scattering does, in general however, not obey electron--hole symmetry.  That the condition is satisfied by Eq.\,\eqref{e.4} is in part a consequence of the neglect of evanescent waves.  These evanescent waves are accounted for in the numerically obtained transmission probability in Fig.\,\ref{f.4}.  Note, for instance, that $T_\tau(0.005\pi/a,0.01\pi/a)\ne T_\tau(-0.005\pi/a,-0.01\pi/a)$.  To understand the lack of electron--hole symmetry in the combined graphene--line defect system, we note that the structure in Fig.\,\ref{f.1}(a) is not bipartite; in particular, the sites participating in the pentagons at the line defect cannot be divided into two types where one type has only nearest neighbors of the other type.

To gain further insight into the numerical calculations and how they lead to Eq.\,\eqref{e.4}, it is useful to perform the transport calculations analytically in the limit of small $q$.  As there is translational symmetry along the line defect, one can focus on those atoms within the primitive cell shown in Fig.\,\ref{f.1}.  The Hamiltonian of the isolated line defect is then
\begin{equation}
	H_D = \gamma\left(
	\begin{array}{cc}
		0 & 1\\
		1 & 0
	\end{array}
	\right).
	\label{e.6}
\end{equation}
Next, we seek a retarded self energy $\Sigma$ that accounts for the coupling of the two line defect atoms to the semi-infinite portion of graphene on each side of the line defect.  As shown in Fig.\,\ref{f.1}, there are two atoms neighboring the line defect on each side.  Expressed in the basis of the atoms parallel to the line defect, the graphene state with valley index $\tau$ is given by
\begin{equation}
	|\tau\rangle = \frac{1}{\sqrt{2}}\left(
	\begin{array}{c}
		1\\
		e^{-2\pi i\tau/3}
	\end{array}
	\right).
	\label{e.7}
\end{equation}
As the basis contains two atoms belonging to the same sublattice, the graphene state above can be folded onto another graphene state with the same wave vector in the full system.  This latter state, however, is evanescent near the Fermi level and can be neglected.  Requiring $\Sigma$ to be retarded fixes the relative phase between the atoms on the line defect and their neighbors, resulting in the relation $\Sigma\langle B|\Phi_\tau\rangle=\gamma\langle A|\Phi_\tau\rangle|\tau\rangle\langle\tau|$, from which one obtains
\begin{equation}
	\Sigma= -\frac{i\gamma}{2}e^{i\tau\alpha}\left(
	\begin{array}{cc}
		1 & e^{2\pi i\tau/3}\\
		e^{-2\pi i\tau/3} & 1
	\end{array}
	\right).
	\label{e.8}
\end{equation}
Equipped with $H_D$ describing the interactions within the line defect and $\Sigma$ describing the coupling to the semi-infinite graphene on each side, one can calculate the retarded Green function on the line defect, $G=\left(\eta\varepsilon I-H_D-2\Sigma\right)^{-1}$, where $I$ is the unit matrix.  To zeroth order in $q$,
\begin{equation}
	G=\frac{-\gamma^{-1}}{1+ie^{i\tau\alpha}}\left(
	\begin{array}{cc}
		ie^{i\tau\alpha} & 1-ie^{i\tau(\alpha+2\pi/3)}\\
		1-ie^{i\tau(\alpha-2\pi/3)} & ie^{i\tau\alpha}
	\end{array}
	\right).
	\label{e.9}
\end{equation}
The probability that the quasi\-particle will transmit through the line defect is given by $T_\tau=\langle\tau|\Gamma G\Gamma G^\dagger|\tau\rangle$, where $\Gamma\equiv i\left(\Sigma-\Sigma^\dagger\right)$.  Inserting Eqs.\,(\ref{e.7}--\ref{e.9}) into this equation, one recovers Eq.\,\eqref{e.4} exactly.

In the initial analysis leading to Eq.\,\eqref{e.4}, an assertion was made that there are two symmetric states at the Fermi level without a node on the line defect.  That claim can now be verified using Eqs.\,(\ref{e.1},\,\ref{e.6},\,\ref{e.8}).  According to Eq.\,\eqref{e.1}, a symmetric state must satisfy $\tau\alpha=\pi/2$.  When this condition is satisfied, one finds that the determinant $\operatorname{det}\left(H_D+2\Sigma\right)=0$, which implies that there are exactly two symmetric states at the Fermi level, one for each valley index $\tau$.  The corresponding eigenstates are $|\Psi_\tau\rangle=|-\tau\rangle$, confirming that the states have no node at the line defect.

Rather than forming isolated Bloch waves, experiments exploiting the valley filter will likely construct more complex wave patterns.  The dimensions of the system should be chosen such that the mean-free path is longer than the distance between the source and the line defect.
For low-energy quasi\-particles, the wavelength $\lambda=2\pi/q$ is much greater than the repeating length $2a$ of the line defect.  As long as this repeating length is also much shorter than any spatial features of the waves, to an excellent approximation the scattering can be treated within ray optics, where rays travel in straight lines and only scatter at the line defect, where they will either transmit with a probability approximately given by Eq.\,\eqref{e.4} or reflect while obeying the law of specular reflection.

The filter can be used to create a valley-polarized beam of electrons or holes.  By probing the current passing through the line defect at a particular angle, one could also measure the valley polarization of the incident quasi\-particles.  Demonstration of these components should significantly accelerate research on graphene valleytronics.

%===========================================================================%

\begin{acknowledgments}
The authors acknowledge support from the U.S. Office of Naval Research, directly and through the U.S. Naval Research Laboratory.  D.G. thanks C.W.J. Beenakker for helpful comments.
\end{acknowledgments}

%===========================================================================%

%

%===========================================================================%

\end{document}